\documentclass[twoside,twocolumn,9pt]{article}
\usepackage{extsizes}
\usepackage[super,sort&compress,comma]{natbib} 
\usepackage[version=3]{mhchem}
\usepackage[left=1.5cm, right=1.5cm, top=1.785cm, bottom=2.0cm]{geometry}
\usepackage{balance}
\usepackage{times,mathptmx}
\usepackage{sectsty}
\usepackage{graphicx} 
\usepackage{lastpage}
\usepackage[format=plain,justification=justified,singlelinecheck=false,font={stretch=1.125,small,sf},labelfont=bf,labelsep=space]{caption}
\usepackage{float}
\usepackage{fancyhdr}
\usepackage{fnpos}
\usepackage[english]{babel}
\addto{\captionsenglish}{%
  
}
\usepackage{array}
\usepackage{droidsans}
\usepackage{charter}
\usepackage[T1]{fontenc}
\usepackage[usenames,dvipsnames]{xcolor}
\usepackage{setspace}
\usepackage[compact]{titlesec}
\usepackage{hyperref}
\usepackage{amsmath,amssymb,bbm}

\usepackage{epstopdf}

\definecolor{cream}{RGB}{222,217,201}

\begin{document}

\pagestyle{fancy}
\thispagestyle{plain}
\fancypagestyle{plain}{

\fancyhead[C]{\includegraphics[width=18.5cm]{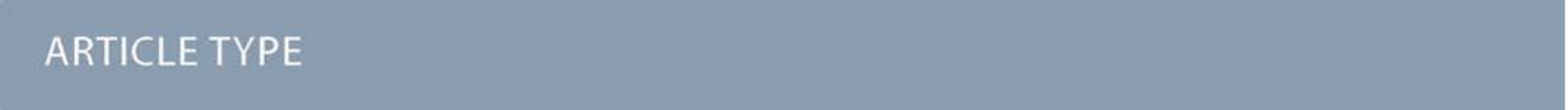}}
\fancyhead[L]{\hspace{0cm}\vspace{1.5cm}\includegraphics[height=30pt]{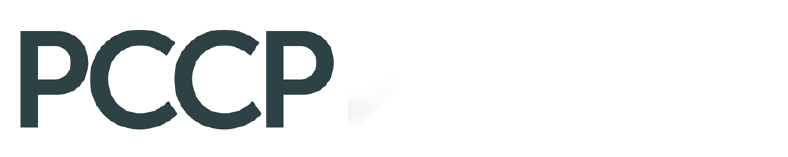}}
\fancyhead[R]{\hspace{0cm}\vspace{1.7cm}\includegraphics[height=55pt]{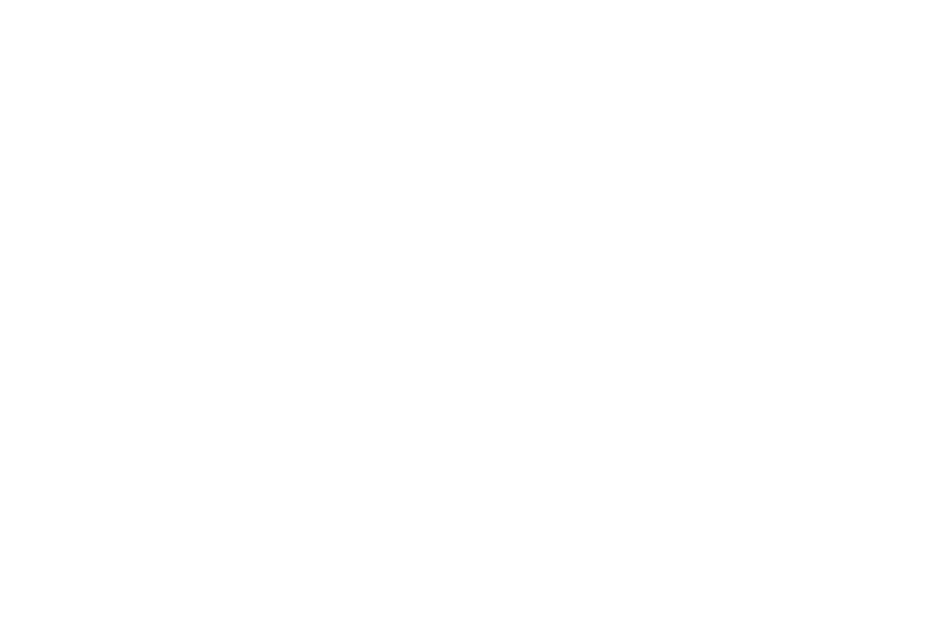}}
\renewcommand{\headrulewidth}{0pt}
}

%
%
\newcommand{\rB}{\mbox{\boldmath$r$}}
\newcommand{\roB}{\mbox{\boldmath$\rho$}}
\newcommand{\RB}{\mbox{\boldmath$R$}}
\newcommand{\uB}{\mbox{\boldmath$u$}}
\newcommand{\thB}{\mbox{\boldmath$\Theta$}}
\newcommand{\ds}{\displaystyle}
\def\pa{{\partial\Omega}}
\def\nmax{n_{\rm max}}
\def\O{\mathcal{O}}
\def\ve{\varepsilon}
\def\R{{\mathbb R}}

\def\S{S}

\makeFNbottom
\makeatletter
\renewcommand\LARGE{\@setfontsize\LARGE{15pt}{17}}
\renewcommand\Large{\@setfontsize\Large{12pt}{14}}
\renewcommand\large{\@setfontsize\large{10pt}{12}}
\renewcommand\footnotesize{\@setfontsize\footnotesize{7pt}{10}}
\makeatother

\renewcommand{\thefootnote}{\fnsymbol{footnote}}
\renewcommand\footnoterule{\vspace*{1pt}%
\color{cream}\hrule width 3.5in height 0.4pt \color{black}\vspace*{5pt}} 
\setcounter{secnumdepth}{5}

\makeatletter 
\renewcommand\@biblabel[1]{#1}            
\renewcommand\@makefntext[1]%
{\noindent\makebox[0pt][r]{\@thefnmark\,}#1}
\makeatother 
\renewcommand{\figurename}{\small{Fig.}~}
\sectionfont{\sffamily\Large}
\subsectionfont{\normalsize}
\subsubsectionfont{\bf}
\setstretch{1.125} 
\setlength{\skip\footins}{0.8cm}
\setlength{\footnotesep}{0.25cm}
\setlength{\jot}{10pt}
\titlespacing*{\section}{0pt}{4pt}{4pt}
\titlespacing*{\subsection}{0pt}{15pt}{1pt}

\fancyfoot{}
\fancyfoot[LO,RE]{\vspace{-7.1pt}\includegraphics[height=9pt]{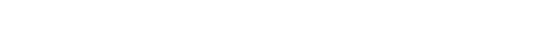}}
\fancyfoot[CO]{\vspace{-7.1pt}\hspace{11.9cm}\includegraphics{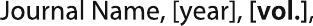}}
\fancyfoot[CE]{\vspace{-7.2pt}\hspace{-13.2cm}\includegraphics{head_foot/RF}}
\fancyfoot[RO]{\footnotesize{\sffamily{1--\pageref{LastPage} ~\textbar  \hspace{2pt}\thepage}}}
\fancyfoot[LE]{\footnotesize{\sffamily{\thepage~\textbar\hspace{4.65cm} 1--\pageref{LastPage}}}}
\fancyhead{}
\renewcommand{\headrulewidth}{0pt} 
\renewcommand{\footrulewidth}{0pt}
\setlength{\arrayrulewidth}{1pt}
\setlength{\columnsep}{6.5mm}
\setlength\bibsep{1pt}

\makeatletter 
\newlength{\figrulesep} 
\setlength{\figrulesep}{0.5\textfloatsep} 

\newcommand{\topfigrule}{\vspace*{-1pt}%
\noindent{\color{cream}\rule[-\figrulesep]{\columnwidth}{1.5pt}} }

\newcommand{\botfigrule}{\vspace*{-2pt}%
\noindent{\color{cream}\rule[\figrulesep]{\columnwidth}{1.5pt}} }

\newcommand{\dblfigrule}{\vspace*{-1pt}%
\noindent{\color{cream}\rule[-\figrulesep]{\textwidth}{1.5pt}} }

\makeatother

\twocolumn[
  \begin{@twocolumnfalse}
\vspace{3cm}
\sffamily
\begin{tabular}{m{4.5cm} p{13.5cm} }

\includegraphics{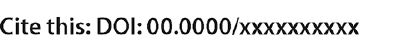} & \noindent\LARGE{\textbf{Diffusion-influenced 
reactions on non-spherical partially absorbing axisymmetric surfaces}} \\
\vspace{0.3cm} & \vspace{0.3cm} \\

 & \noindent\large{Francesco Piazza,\textit{$^{a}$} Denis Grebenkov,\textit{$^{b}$}} \\

\includegraphics{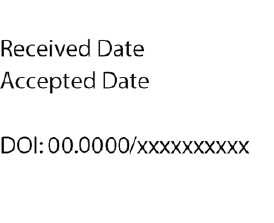} & \noindent\normalsize{
The calculation of the diffusion-controlled reaction rate for
partially absorbing, non-spherical boundaries presents a formidable
problem of broad relevance.  In this paper we take the reference case
of a spherical boundary and work out a perturbative approach to get a
simple analytical formula for the first-order correction to the
diffusive flux onto a non-spherical partially absorbing surface of
revolution.  To assess the range of validity of this formula, we
derive exact and approximate expressions for the reaction rate in the
case of partially absorbing prolate and oblate spheroids.  We also
present numerical solutions by a finite-element method that extend the
validity analysis beyond spheroidal shapes.  Our perturbative solution
provides a handy way to quantify the effect of non-sphericity on the
rate of capture in the general case of partial surface reactivity.}
\\

\end{tabular}

 \end{@twocolumnfalse} \vspace{0.6cm}

  ]

\renewcommand*\rmdefault{bch}\normalfont\upshape
\rmfamily
\section*{}
\vspace{-1cm}


\footnotetext{\textit{$^{a}$~Centre de Biophysique Mol\'eculaire (CBM) CNRS UPR4301 $\&$ Universit\'e d'Orl\'eans, Orl\'eans 45071, France; Tel: +33 238 255653; E-mail: Francesco.Piazza@cnrs-orleans.fr}}
\footnotetext{\textit{$^{b}$~Laboratory of Condensed Matter Physics, CNRS -- Ecole Polytechnique, IP Paris, F-91128 Palaiseau, France. }}





\section{Introduction}

Reaction-diffusion processes govern the behavior of complex systems in
many contexts ranging from chemistry and
nanosciences~\cite{Palazzo:2014aa} to
bio-engineering~\cite{rice1985,Welsch:2009aa,Galanti:2016ac,Roa:2018aa}.
They are also key in biology by controlling enzyme
catalysis~\cite{Gopich:2013aa,Welsch:2009aa}, antigen-antibody
encounter~\cite{Bongini:2007az,Piazza:2005vn}, ligand-receptor
binding~\cite{Berg77,Shoup81,Shoup:aa,Zwanzig90,Zwanzig91,Lauffenburger},
fluorescence quenching~\cite{Tachiya2007}, cellular nutrient
uptake~\cite{berg1993,karp1996,kiorboe2008,Sozza:2018aa}, and oxygen
uptake in lungs and placenta \cite{Sapoval02,Grebenkov05,Serov16}.\\
\indent In many cases, diffusion relaxation times are fast with respect to typical 
time scales of interest. For example, the local concentration field of
ligand molecules around a membrane receptor equilibrates more swiftly
than lateral diffusion of the latter within the membrane and more
rapidly than (or as fast as) large-scale conformational changes of the
protein.  Along the same lines, the concentration of a small substrate
molecule equilibrates fast around a large core-shell
nanoreactor~\cite{Roa:2018aa} whose hydrogel shell carries a number of
immobilized metal nanocatalysts.  As a consequence, the problem can
often be reduced to solving the stationary diffusion (i.e. Laplace)
equation~\cite{rice1985} in a non-equilibrium setting
(i.e. corresponding to a steady non-zero flux of molecules from the
bulk to the surface).  In the simplest case, one is led to computing
the total molecular flux to a reactive surface, modeling for example
the uptake of nutrient molecules by a colony of algae or the binding
of a ligand onto the receptor-covered cell surface.\\
\indent 
In more mathematical terms, we consider the pseudo first-order
contact reaction
\begin{equation}
\label{e:reaction}
\ce{B + S ->[k] S + P}
\end{equation}
between point-like molecules $B$ that diffuse in solution with
diffusion coefficient $D$ and static reactive surfaces $S$ that
catalyze the transformation of $B$ into some inert product $P$.  When
the characteristic relaxation time for diffusive processes is small
enough, the kinetics is essentially controlled by the steady-state
rate $k$~\cite{rice1985}.  If $B$ and $S$ species are dilute enough
and the concentration of $S$ is much smaller than the bulk
concentration of $B$, $c_\infty$, the rate $k$ can be obtained by
solving the Laplace equation with appropriate boundary conditions (BC)
enforced on the surface $\S$ and computing the total steady-state
diffusive flux of $B$ molecules into an isolated
$\S$~\cite{rice1985,Piazza:2013aa,Galanti16,Grebenkov19b}.  When
$\S$ is a sphere of radius $R$, this leads to the well-known
Smoluchowski rate~\cite{Smoluchowski:1916fk,Smoluchowski:1917aa},
\begin{equation}
\label{e:kS}
k_S = 4\pi D R c_\infty
\end{equation}
which expresses the very blueprint of diffusion to capture: the number
of particles absorbed at the sink surface scales linearly with the
size of its surface.\\
\indent In real-life applications, reactive boundaries tend to
be more complex than perfectly absorbing spheres.  On one hand, many
interesting applications feature {\em partially} reactive surfaces
(see a recent overview in \cite{Grebenkov19}).  The partial reactivity
can describe, e.g., the need for a molecule to overcome an energy
activation barrier to react \cite{Weiss86}, or heterogeneous
distribution of reactive patches on the otherwise inert boundary when
small swiftly diffusing molecules can only react at
specific parts of the boundary
\cite{Berezhkovskii04,Berezhkovskii06,Muratov08,Dagdug16,Lindsay17,Bernoff18a,Bernoff18b,Grebenkov19c}.
This is the case for example of a more realistic treatment of a cell
surface
\cite{Berg77,Shoup81,Shoup:aa,Zwanzig90,Zwanzig91}, where receptor
molecules only constitute small absorbing patches (or clusters) on an
otherwise inert surface%
\footnote{We refer in this setting to the
inability of the diffusing ligand to establish bonds at generic spots
on the surface, as opposed to specific high-affinity bonds at
locations occupied receptors.  This does not exclude that the surface
might be interacting non-specifically (and less strongly) with the
ligand at generic locations.}.  Partial reactivity can also describe
stochastic gating of exchange channels (like aquaporins) on the plasma
membrane \cite{Benichou00,Reingruber09,Lawley15,Bressloff17}, partial
recombination \cite{Sano79}, reactions on micelles \cite{Sano81}, and
transfer across semi-permeable membranes
\cite{Sapoval94,Grebenkov06}.  On the other hand, in many applications
one has to deal with {\em non-spherical}
boundaries~\cite{Piazza:2005vn}.  This in general makes the problem of
determining the diffusive flux very hard already in the simpler case
of a perfect sink \cite{Traytak:2018ab,Grebenkov:2018aa}, the case of
partially absorbing non-spherical surfaces being in general a rather
formidable task.  This is the case for example of binding to a
protein, featuring a few specific binding pockets (local absorbing
patches) on an overall reflecting surface that is most of the times
non-spherical. \\
\indent 
In this paper, we tackle the hard problem of computing the flux to
non-spherical surfaces perturbatively.  Our calculations lead to a
handy, closed-form expression for the first-order correction to the
Smoluchowski rate for arbitrary, partially absorbing axially symmetric
perturbations of a spherical boundary.  The paper is organized as
follows.  In section \ref{sec:solution} we state the problem and work
out the perturbative solution.  In section \ref{sec:results} we report
a detailed analysis of the validity regime of the first-order
correction.  For this purpose, the diffusive flux to partially
reactive prolate and oblate spheroids is computed.  We also present
numerical computations for more sophisticated surfaces.  Finally, we
wrap our results and summarize in section \ref{sec:summary}.

%
\section{Methodology}
\label{sec:solution}

%
\begin{figure}[t!]
\centering
\includegraphics[width=5.5truecm]{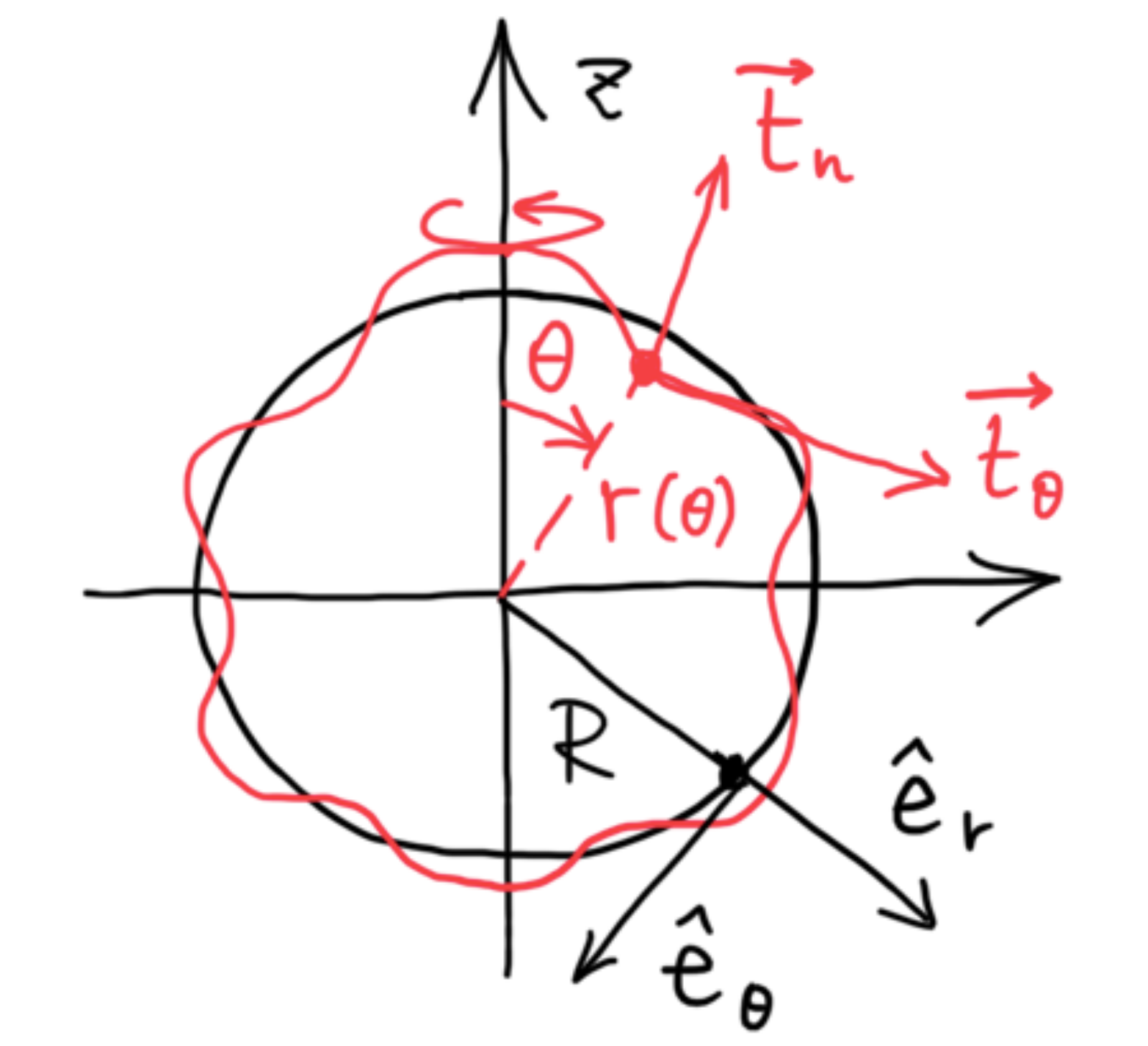}
\caption{
Scheme of a surface of revolution (red) obtained as a perturbation of
the sphere $\S_R$ (black), $r(\theta) = R + \delta
\,f(\theta)$ with $\delta \ll R$.}
\label{f:scheme}
\end{figure}

\noindent 
Let us consider the surface of revolution $\partial \Omega$,
parameterized in spherical coordinates $(r,\theta,\phi)$ as
(Fig.~\ref{f:scheme})
\begin{eqnarray}
\label{e:Omf}
x &=& r(\theta) \cos\phi \sin\theta \nonumber \\
y &=& r(\theta) \sin\phi \sin\theta \\
z &=& r(\theta) \cos\theta \nonumber
\end{eqnarray}
with $\theta\in[0,\pi]$, $\phi\in[0,2\pi)$ and
\begin{equation}
\label{e:rtheta}
r(\theta) = R + \delta \, f(\theta), \qquad \delta \ll R
\end{equation}
The idea is to consider an arbitrary small modulation of the sphere
$\S_R$ of radius $R$, as specified by a given function
$f(\theta)$.  We aim to compute the steady-state diffusive flux of
point-like $B$ particles into the partially absorbing surface
$\partial \Omega$.  As put forward by Collins and
Kimball~\cite{Collins:1949aa}, this situation can be accommodated for
within the framework of a boundary value problem by endowing the
reactive surface with a constant intrinsic reactivity $\kappa$ (in
units m/s).  Whenever a diffusing $B$ particle hits the surface, the
reaction~\eqref{e:reaction} occurs with a probability proportional to
$\kappa$ \cite{Filoche99,Grebenkov03,Grebenkov06a}.  From a
mathematical standpoint, one needs to compute the steady-state
concentration field of $B$ particles, $c(\vec{r})$, satisfying the
following boundary value problem:
\begin{subequations}\label{e:SmolBP}
\begin{align}
&\nabla^2 c = 0 \qquad  \vec{r} \in \mathbb{R}^3 \setminus T    \label{e:SmolBP1}\\
&D \nabla c \cdot \hat n|_{\partial \Omega} = \kappa \, c|_{\partial \Omega} \label{e:SmolBP2}\\
&\lim_{||\vec{r}||\to\infty} c = c_\infty                                    \label{e:SmolBP3}
\end{align}
\end{subequations}
where $T$ is the interior of the surface $\partial\Omega$.
The Robin boundary condition~\eqref{e:SmolBP2} means that the
diffusive flux density should equal the reactive flux density at each
point of the surface.  In this paper, we take the convention of
considering the normal pointing away from the boundary into the bulk;
in other words, we consider the inward normal vector, which can be
computed as the cross product of the two local tangent vectors (see
Fig.~\ref{f:scheme}):
\begin{equation}
\label{e:tn}
\hat{n} = \frac{\vec{t}_n}{||\vec{t}_n||} = 
          \frac{\vec{t}_\theta \wedge \vec{t}_\phi}{|| \vec{t}_\theta \wedge \vec{t}_\phi ||}
\end{equation}
In the limit $\kappa\to\infty$ the boundary becomes a sink (Dirichlet
BC), i.e. $c|_{\partial \Omega}=0$, while in the opposite limit
$\kappa\to 0$ (von Neumann BC), the surface becomes perfectly
reflecting and no reaction occurs.  The regularity
condition~\eqref{e:SmolBP3} means that the concentration field of $B$
molecules far from the boundary is constant and equal to the bulk
concentration.  This ensures that after a relaxation transient period,
the system settles in a nonequilibrium steady state with a net
diffusive flux onto the reactive boundary.  Accordingly, the reaction
rate can be computed as the total diffusive flux into the surface,
i.e.
\begin{equation}
\label{e:rate}
k = -\int_{\partial \Omega} \vec{J} \cdot \hat n \, dS 
= 2\pi D \int_0^\pi \nabla c\cdot \hat{n} \,  \, d\theta
\end{equation}
where $\vec{J} = -D\nabla c$ is the flux density.  It is
expedient to introduce non-dimensional variables, $\vec{\rho}=\vec{r}/R$,
$u(\vec{\rho}) = 1 - c(\vec{\rho} R)/c_\infty$, $\epsilon = \delta/R\ll 1$, and $h
= \kappa R/D$. The problem to be solved reads then
\begin{subequations}\label{e:SmolBPx}
\begin{align}
&\nabla^2 u = 0 \qquad \vec{\rho} \in  \mathbb{R}^3 \setminus T'     \label{e:SmolBPx1}\\
&\left. \biggl(
   \nabla u \cdot \hat{n} + h(1-u)
 \biggr)\right|_{\partial \Omega'} = 0                                       \label{e:SmolBPx2}\\
&\lim_{\rho\to\infty} u = 0                                             \label{e:SmolBPx3}
\end{align}
\end{subequations}
where $\rho=||\vec{\rho}||$, and prime emphasizes that the surface
$\partial\Omega$ and its interior $T$ are also rescaled. \\
\indent The above problem can be solved perturbatively by looking for a
solution in the form of a perturbative expansion in powers of
$\epsilon$.  Furthermore, the symmetry of the problem requires the
solution to be axially symmetric.  To the first order, we set then
\begin{equation}
\label{e:uPE}
u(\rho,\theta) = u_0(\rho) + \epsilon \, u_1(\rho,\theta) +\mathcal{O}(\epsilon^2)
\end{equation}
where $u_0$ and $u_1$ are both harmonic functions that can be
expressed in general as
\begin{subequations}
\begin{align}
& u_0(\rho) = \frac{A_0}{\rho}   \label{e:u0u11} \\
& u_1(\rho,\theta) = \sum_{\ell=0}^\infty B_\ell \,\rho^{-(\ell+1)}P_\ell(\cos\theta) \label{e:u0u12}
\end{align}
\end{subequations}
Here $P_\ell(\cos\theta)$ denotes the Legendre polynomial of order
$\ell$ and $\{A_0,B_0,B_1,\dots\}$ is an infinite set of constants
that are fixed by the boundary condition~\eqref{e:SmolBPx2}.\\
\indent In order to impose the BC~\eqref{e:SmolBPx2}, we need to work out 
the gradient $\nabla u$ and the concentration field $u$ on the
surface $\partial \Omega'$, as well as the components of the inward
normal vector $\vec{t}_n$.  From the definition~\eqref{e:tn}, it is
straightforward to show that
\begin{subequations}
\begin{align}
\vec{t}_n &= \bigl( 
              [1 + 2\epsilon f(\theta)] \,\hat{e}_r - \epsilon f^\prime({\theta})\,\hat{e}_\theta
             \bigr) \sin\theta     + \mathcal{O}(\epsilon^2) \label{e:normalgrad1} \\
u|_{\partial \Omega'} &= u_0(1) + \epsilon 
                              \left[
                                      u_1 + f(\theta) \frac{d u_0}{d \rho}
                              \right]_{\rho=1} + \mathcal{O}(\epsilon^2) \label{e:normalgrad0} \\                     
\nabla u|_{\partial \Omega'} &= \left[
              \frac{d u_0}{d \rho} + \epsilon
              \left( 
                    f(\theta) \frac{d^2 u_0}{d \rho^2} + 
                    \frac{\partial u_1}{\partial \rho}
              \right)
            \right]_{\rho=1} \hat{e}_r +  \epsilon \frac{\partial u_1}{\partial \theta} \Big|_{\rho=1}
                                  \hat{e}_\theta + \mathcal{O}(\epsilon^2) \label{e:normalgrad2}
\end{align}          
\end{subequations}
where $\hat{e}_r$ and $\hat{e}_\theta$ are the radial and polar
tangent unit vectors relative to the sphere $\S_R$ (see
Fig.~\ref{f:scheme}). With the help of Eqs.~\eqref{e:normalgrad1}
and~\eqref{e:normalgrad2}, the boundary conditions~\eqref{e:SmolBPx2}
and~\eqref{e:SmolBPx3} give
\begin{subequations}
\begin{align}
&A_0 = \frac{h}{1+h} \label{e:A0}\\
&\sum_{\ell=0}^\infty  (\ell+h+1)B_\ell P_\ell(\cos\theta) = \frac{h(2+h)}{1+h} f(\theta)\label{e:Bleq}
\end{align}
\end{subequations}
Multiplying Eq.~\eqref{e:Bleq} by $P_m(\cos\theta)\sin\theta$, integrating and 
recalling the orthonormality of Legendre polynomials,
\begin{equation}
\label{e:LegP}
\int_{-1}^1 P_{\ell}(\mu) P_{{\ell^\prime}}(\mu) \, d\mu = \frac{2}{2\ell+1} \delta_{\ell\ell^\prime}
\end{equation}
one is immediately led to  
\begin{equation}
\label{e:Bl}
B_\ell = \frac{(2\ell+1)(2+h)h}{2(\ell + 1 + h)(1+h)} \int_0^\pi P_\ell(\cos\theta) f(\theta) \sin\theta \,d\theta
\end{equation}
%

\section{Results and discussion}
\label{sec:results}

The rate should be computed from Eq.~\eqref{e:rate}.  Taking into
account Eqs.~\eqref{e:normalgrad1},~\eqref{e:normalgrad2} 
and~\eqref{e:LegP}, we get
\begin{eqnarray}
\label{e:ratef}
\frac{k}{k_S} &=& \frac{h}{1+h} + \epsilon B_0 + \mathcal{O}(\epsilon^2)  \nonumber\\
              &=& \frac{h}{1+h}\left[
              1 + \left( 
                         \frac{2+h}{1+h} 
                  \right)
                  \epsilon B^\infty_0 
              \right]
               + \mathcal{O}(\epsilon^2)
\end{eqnarray}
where 
\begin{equation}
\label{e:B0}
B_0^\infty = \frac{1}{2} \int_0^\pi f(\theta) \sin\theta \,d\theta
\end{equation}
Formula~\eqref{e:ratef} is the main result of this paper, describing
the first-order correction to the Smoluchowski rate for a
non-spherical, axisymmetric partially reactive boundary.  In the limit
of a perfect sink $h\to\infty$, this becomes
\begin{equation}
\label{e:ratesink}
\frac{k}{k_S} = 1 + \epsilon B^\infty_0 + \mathcal{O}(\epsilon^2) 
\end{equation}
It is instructive to observe that the flux to the partially absorbing
boundary and the flux to the sink are not related to the first order
in $\epsilon$ through the usual additivity prescription for
independent probabilities (per unit time):
\begin{eqnarray}
\label{e:invk}
\frac{1}{k} &\simeq& \frac{1}{k_r} + \frac{1}{k_S\left[ 
                                                   1 + \left(\frac{2+h}{h} \right) 
                                                   \epsilon  B_0^\infty 
                                                 \right]} \nonumber\\
            &\neq& \frac{1}{k_r} + \frac{1}{k_S(1 + \epsilon B_0^\infty )}
\end{eqnarray}
where $k_r = h k_S = 4\pi R^2 \kappa$ has dimensions of a rate constant and 
gauges the uniform chemical activity of the boundary.  
Eqs.~\eqref{e:invk} and \eqref{e:ratef} coincide
to the first order in $\epsilon$ only for highly reactive surfaces
($h\gg1$). It is worthwhile to mention that the information carried by the intrinsic
reaction rate $k_r$ it is also known as Damk\"{o}hler number Da, 
\begin{equation}
\label{e:Da}
\textrm{Da} = \frac{\kappa R}{D} = \frac{k_r}{k_S} = h
\end{equation}
which quantifies to what extent the overall reaction rate is limited by diffusion.

\subsection{Validity range of the perturbative calculation}
\label{sec:validity}

\begin{figure}[t!]
\centering
\includegraphics[width=\columnwidth]{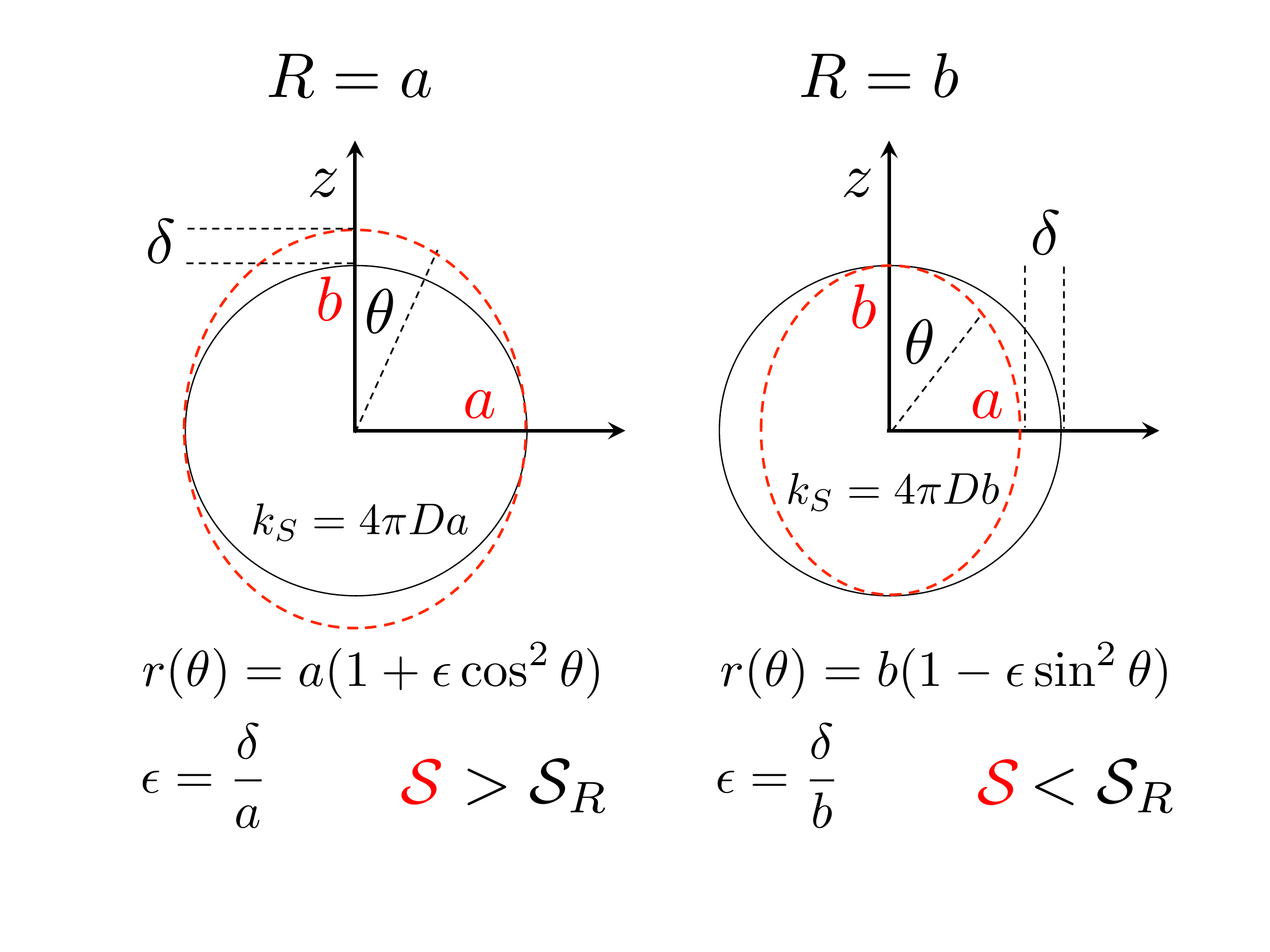}
\caption
{A prolate spheroid can be obtained as a perturbation of a spherical
surface either by increasing or decreasing the surface of the latter.
The first-order correction to the Smoluchowski rate corresponding to
the sphere will be positive in the first case and negative in the
second.  A similar analysis applies to oblate spheroidal perturbations
of a sphere.}
\label{f:spheroid}
\end{figure}

\noindent 
In order to determine the validity range of our results, we consider
the case of spheroidal perturbations (i.e. ellipsoids of rotation). As
an example, Fig.~\ref{f:spheroid} illustrates a prolate spheroidal
perturbation of the form $(a,a,b)$ with $a<b$ that is obtained from a
sphere of radius $R$ either by extension along the $z$-axis into
$(R,R,R+\delta)$, or by compression at the equator into
$(R-\delta,R-\delta,R)$.  In the compression case, the surface area of
the resulting spheroid is reduced as compared to that of the sphere
$\S_R$, whereas it increases in the extension case.
Accordingly, the first-order correction to the Smoluchowski rate is
expected to be negative in the compression case and positive in the
extension case.  A similar reasoning applies to the case of oblate
spheroidal perturbations of the form $(b,b,a)$ with $a<b$.\\
%

\subsection{Perfectly absorbing spheroids}

In order to illustrate our calculations, we first treat the case of
spheroidal perturbations of a spherical sink.  Let us consider a
perfectly absorbing sphere (i.e. $h\to\infty$) with radius $R=b$.  With reference to
Fig.~\ref{f:spheroid} (right), the surface area of a prolate spheroid
with axes $(a,a,b)$, with $a = R-\delta$, reads
\begin{eqnarray}  \nonumber
\S &=& 
\frac{\S_R}{2} \left[
								\frac{1}{(1+\epsilon)^2} + \frac{1}{\sqrt{\epsilon^2+2\epsilon}}
								\arcsin \left(
								                \frac{\sqrt{\epsilon^2+2\epsilon}}{1+\epsilon}
								        \right)
						\right] \\
\label{e:Surfspheroideps}
&=& \S_R \left( 1 - \frac{4}{3}  \epsilon  \right) + \mathcal{O}(\epsilon^2),
\end{eqnarray}
with $\epsilon = \delta/R = 1 - a/b$.  An equivalent analysis can be
performed in the case of oblate spheroids.  Overall, it is not
difficult to see (cf. also Appendix \ref{sec:prolate}) that the
perturbation modulating function $f(\theta)$ for spheroidal
perturbations of the compression type reads
\begin{equation}
\label{e:fthprob}
f(\theta) = 
\begin{cases}
-\sin^2\theta &  \text{prolate}  \\
-\cos^2\theta &  \text{oblate}  \\
\end{cases}
\end{equation}
while for perturbation of the extension type, one has
\begin{equation}
\label{e:fthprobstr}
f(\theta) = 
\begin{cases}
\cos^2\theta &  \text{prolate}  \\
\sin^2\theta &  \text{oblate}  \\
\end{cases}
\end{equation}
Taking into account expressions~\eqref{e:fthprob}, the
prescription~\eqref{e:ratesink} gives immediately the negative
corrections for the compression case,
\begin{equation}
\label{e:ratesspheps1}
\frac{k}{k_S} = 
\begin{cases}
1 - \frac{2}{3}\epsilon + \mathcal{O}(\epsilon^2) & \text{prolate}  \\
1 - \frac{1}{3}\epsilon + \mathcal{O}(\epsilon^2)& \text{oblate}  \\
\end{cases}
\end{equation}
where $k_S = 4\pi D b c_\infty$.  The
expressions~\eqref{e:ratesspheps1} are in agreement with the known
exact results (see
\cite{Berezhkovskii:2007aa,Traytak:2018ab,Grebenkov:2018aa} and
Appendices).  The corrections for the extension case can be computed
easily from Eqs.~\eqref{e:fthprobstr},
\begin{equation}
\label{e:ratesspheps2}
\frac{k}{k_S} = 
\begin{cases}
1 + \frac{1}{3}\epsilon + \mathcal{O}(\epsilon^2) & \text{prolate}  \\
1 + \frac{2}{3}\epsilon + \mathcal{O}(\epsilon^2)& \text{oblate}  \\
\end{cases}
\end{equation}
where the unperturbed rate is now $k_S = 4\pi D a c_\infty$ (see also
Fig.~\ref{f:scheme} left panel)%
\footnote{These results can also be
obtained by a simple rescaling argument.  Setting $b = a/(1-\epsilon)$
into $k_S = 4\pi D b c_\infty$ in Eq.~\eqref{e:ratesspheps1}, one
immediately recovers expressions~\eqref{e:ratesspheps2} from $k/(4\pi
D a c_\infty)= k/[4\pi D b c_\infty(1-\epsilon)]$.}.\\
\indent These results have a very simple interpretation.  Indeed, they 
correspond to the Smoluchowski rates into the respective {\em
equivalent spheres} $\S_{R_e}$, that is, spheres with the
same surface (to the first order in $\epsilon$).  From
Eqs.~\eqref{e:Surfspheroideps} it can be seen immediately that $R_e =
R \sqrt{1- 4\epsilon/3}$ (prolate) and $R_e = R \sqrt{1-2\epsilon/3}$
(oblate).  This suggests that in general the diffusive flux into a
non-spherical boundary can be approximated by the Smoluchowski rate
corresponding to its equivalent sphere to the first order in the
difference of the relevant linear dimensions of the two surfaces.
This agrees with the results of previous calculations by Berezhkovskii
and Barzykin~\cite{Berezhkovskii:2007aa}.\\
\indent Our analysis shows that perturbing a sphere
into a spheroid with given aspect ratio (to the first order in
$\epsilon$) does not result in symmetric corrections to the
Smoluchowski rate of the sphere, depending on whether the latter is
compressed in the equatorial plane or elongated along the polar axis.
More precisely, reducing a spherical surface to a prolate spheroid
implies losing twice as much flux than reducing it to an oblate one.
Conversely, increasing a spherical surface to a prolate spheroid
implies gaining half as much flux than increasing it to an oblate
one.\\

%
\begin{figure*}[t!]
\begin{center}
\includegraphics[width=18truecm]{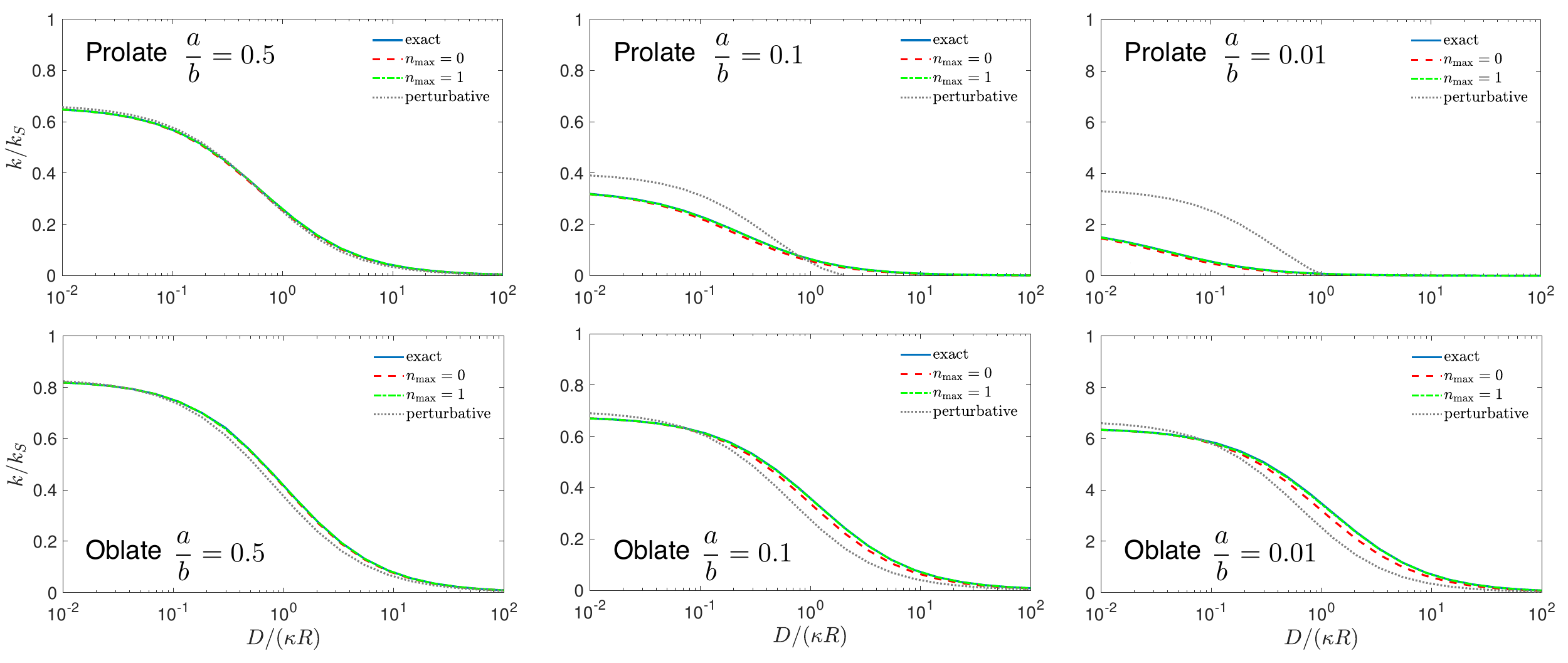}
\end{center}
\caption{
Total diffusive flux to partially absorbing spheroids as a function of the 
inverse Damk\"{o}hler number 
$D/(\kappa b)$ (rescaled by $k_S = 4\pi D b c_\infty$) for different
aspect ratios.  Solid lines show the numerical solution with the
truncation order $n_{\rm max} = 5$ (referred to as ``exact solution'',
see Appendices), while dashed and dashed-dotted lines represent the
approximate solutions corresponding $n_{\rm max} = 0$
(i.e. Eq.~\eqref{e:ratestrunc}) and $n_{\rm max} = 1$, respectively.
Dotted lines show the perturbative solutions~\eqref{e:ratef}.
Top panels: prolate spheroids $(a,a,b)$; Bottom panels: oblate
spheroids $(b,b,a)$.}
\label{fig:prol_obl_k}
\end{figure*}
%

\subsection{Partially absorbing spheroids}

\noindent

While the diffusive flux onto partially absorbing prolate and oblate
spheroids can be obtained by solving infinite-dimensional systems of
linear equations (see Appendices), its exact closed-form expressions
are not available.  However, it is possible to obtain approximate
analytical expressions that turn out to be remarkably accurate for a
wide range of aspect ratios and reactivities.  By truncating the
infinite set of linear equations to the zeroth order (see Appendices
for the detailed calculations), we obtain in the case of compressed
spheroids
\begin{equation}
\label{e:ratestrunc}
\frac{k}{k_S} = 
\begin{cases}  \displaystyle
\frac{a_E}{b}
\biggl(\frac12 \ln \biggl(\frac{1+a_E/b}{1-a_E/b}\biggr) + \frac{D}{\kappa b} 
\frac{\arcsin(a_E/b)}{(1 - a_E^2/b^2)^{1/2}}\biggr)^{-1}   \hskip 4mm \text{prolate}  \\
\displaystyle
\frac{a_E/b}{\arcsin(a_E/b)} \, \frac{1}{1 + D/(\kappa b)}  \hskip 33mm \text{oblate}  \\
\end{cases}
\end{equation}
where $k_S = 4\pi D b c_\infty$ and $a_E = \sqrt{b^2 - a^2} =
b\sqrt{2\epsilon-\epsilon^2}$.  These expressions are compared in
Fig.~\ref{fig:prol_obl_k} to the exact results (numerical solutions of
the linear system to a fixed accuracy, cf Appendices) and to the
perturbative expressions~\eqref{e:ratesspheps1}.  One can see that the
zeroth-order approximations~\eqref{e:ratestrunc} are barely
distinguishable from the exact solutions over a very broad range of
aspect ratios.  More precisely, these expressions still yield accurate
predictions of the total flux even for $a\ll b$, i.e. very elongated
prolate spheroids and very flat oblate spheroids.  Moreover, they are
valid over the whole range of reactivities, from perfectly
absorbing surfaces ($\kappa\to\infty$) to nearly reflecting boundaries
($\kappa\to 0$).\\
\indent Figure~\ref{fig:prol_obl_k} also illustrates clearly the range of 
validity of our perturbative calculations.  The simple first-order
perturbative corrections that one may compute easily from
Eq.~\eqref{e:ratef} provides a remarkably reliable estimate of the
rate for values of $\epsilon =1-a/b$ as large as 0.5.  In the case of
oblate spheroids, Fig.~\ref{fig:prol_obl_k} shows that the
perturbative solution is an utterly reasonable approximation even for
disk-like spheroids ($a = 0$).  Conversely, the flux onto thinner and
thinner prolate spheroids become smaller and smaller, and consequently
the perturbative solution fails in the limit $a\to 0$ (right-most
upper panel in Fig.~\ref{fig:prol_obl_k}).

\subsection{Beyond spheroidal shapes}

\begin{figure}[ht]
\begin{center}
\includegraphics[width=\columnwidth]{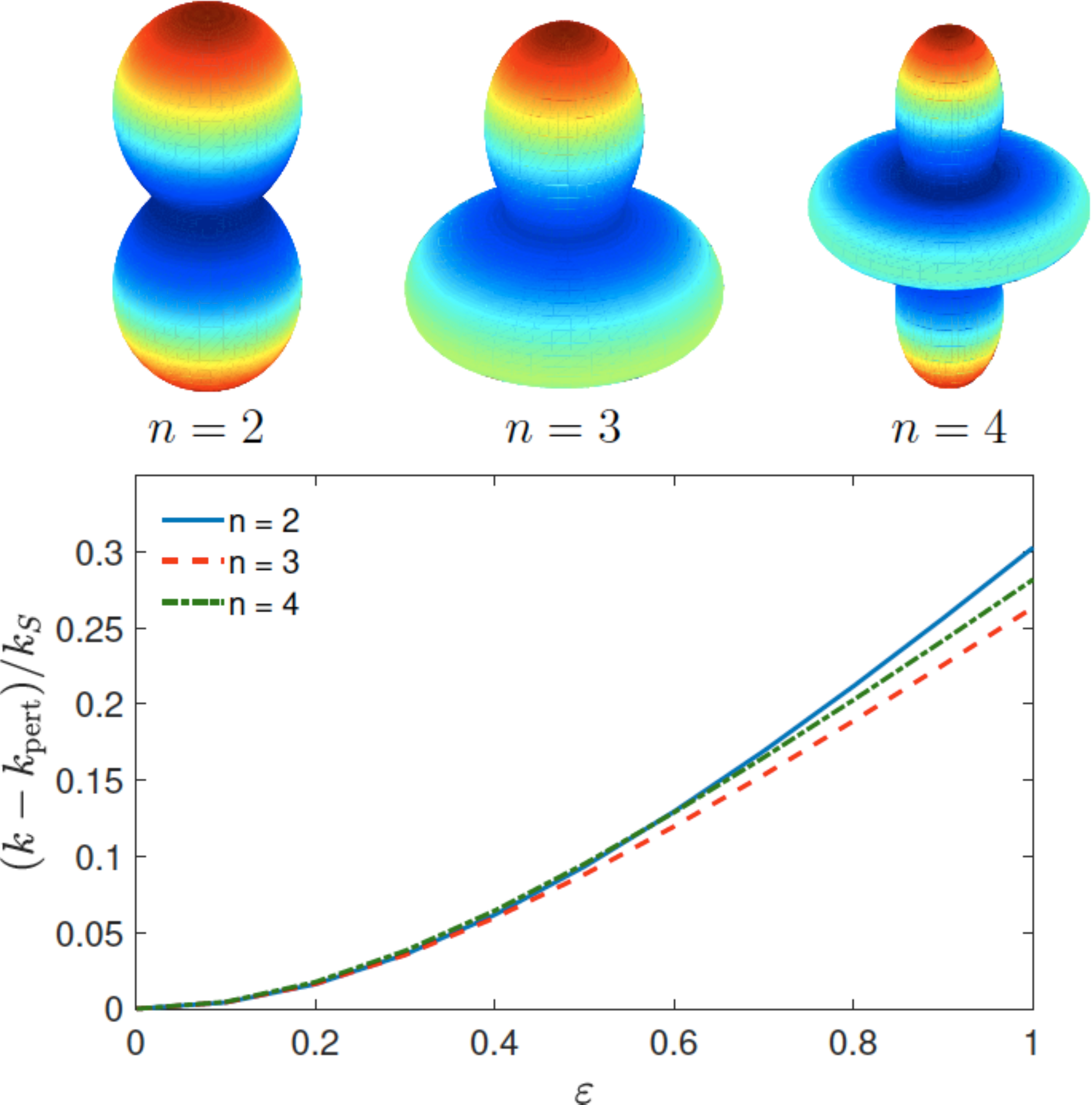}
\end{center}
\caption{
(Top) Three axisymmetric surfaces defined by
Eq.~\eqref{eq:r_examples1}, with $\ve = 1$ and $n = 2,3,4$. (Bottom)
Difference between the total flux $k$ to these perfectly reactive
surfaces ($h\to \infty$) and its first-order approximation $k_{\rm pert}= k_S$, 
rescaled by the Smoluchowski rate 
$k_S = 4\pi D c_\infty R$.  The finite-element computation was
performed in Matlab PDEtool, with the outer boundary at $R_\alpha = 10
R$ and the maximal mesh size $\eta_{\rm max}$ of $0.05$.  Increasing
$R_\alpha$ or decreasing $\eta_{\rm max}$ did not affect the computed
flux, confirming the high numerical accuracy (not shown).}
\label{fig:examples1}
\end{figure}
\begin{figure}[ht]
\begin{center}
\includegraphics[width=\columnwidth]{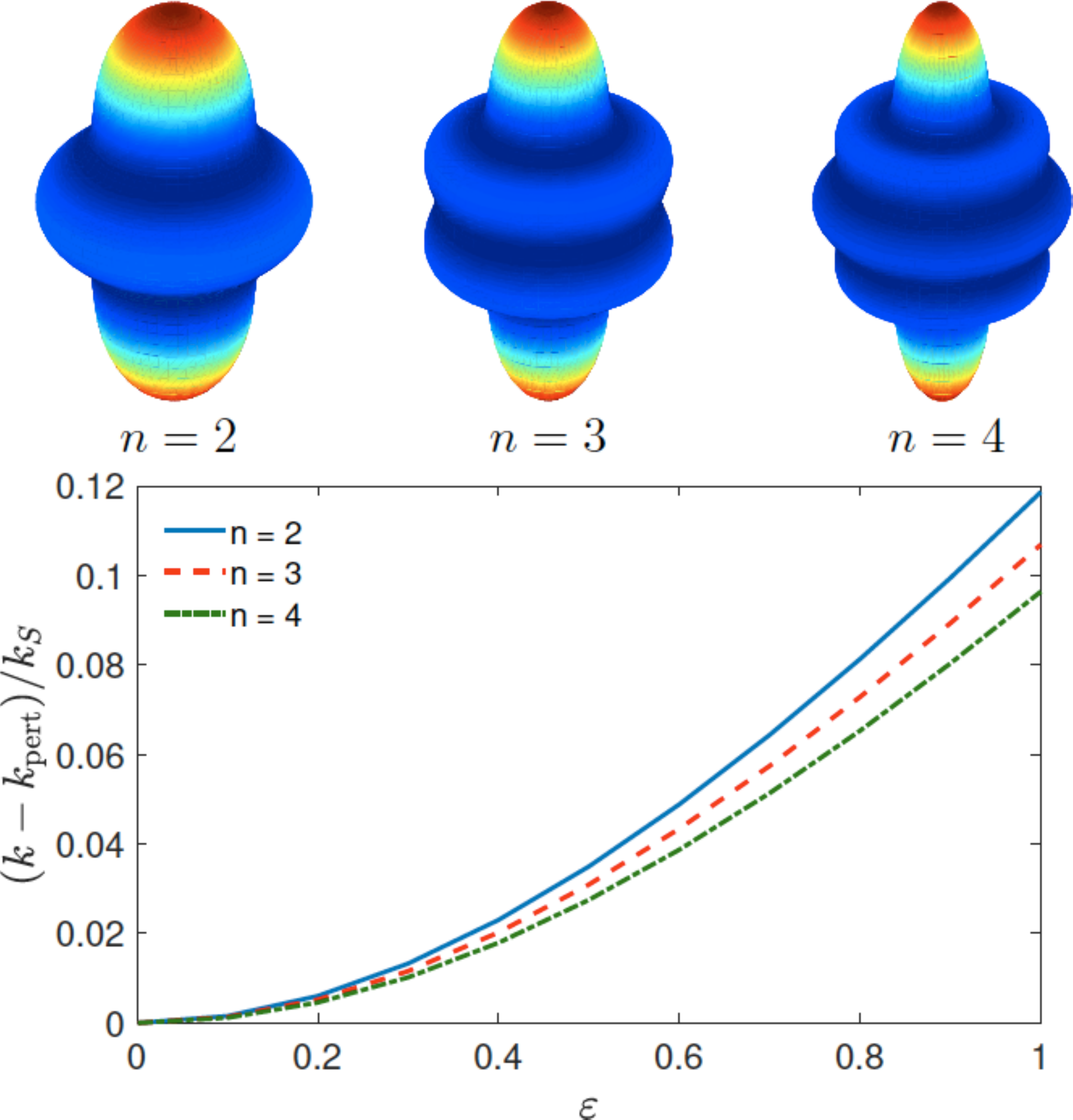}
\end{center}
\caption{
(Top) Three axisymmetric surfaces defined by
Eq.~\eqref{eq:r_examples2}, with $\ve = 1$ and $n = 2,3,4$. (Bottom)
Difference between the total flux $k$ to these perfectly reactive
surfaces ($h\to \infty$) and its first-order approximation $k_{\rm
pert}$ from Eq.~\eqref{eq:k_pert2}), rescaled by the Smoluchowski rate 
$k_S = 4\pi D c_\infty R$.  The finite-element computation was
performed in Matlab PDEtool, with the outer boundary at $R_\alpha = 10
R$ and the maximal mesh size $\eta_{\rm max}$ of $0.05$.  Increasing
$R_\alpha$ or decreasing $\eta_{\rm max}$ did not affect the computed
flux, confirming the high numerical accuracy (not shown).}
\label{fig:examples2}
\end{figure}

%
In order to assess the accuracy of our perturbative approximation
beyond spheroidal shapes, we solved the boundary value problem~\eqref{e:SmolBPx} 
through a finite-element method implemented in Matlab PDEtool 
(see Appendix~\ref{sec:A_FEM} for details).  We then compute
numerically the total flux $k$ in the case of 
perfectly absorbing boundary conditions ($h = \kappa R/D \to \infty$), 
which is compared to our first-order approximation
\begin{equation}  \label{eq:k_pert}
k_{\rm pert} = k_S \biggl(1 +  \epsilon B_0^\infty\biggr) 
\end{equation}
where  $B_0^\infty$ is determined uniquely by the
shape of the surface via Eq.~\eqref{e:B0}.  Even though this numerical
scheme is applicable to any axisymmetric surface determined by a given
function $r(\theta)$, we focus on two families of surfaces.  The first
family is determined by Legendre polynomials $P_n(z)$,
\begin{equation}  
\label{eq:r_examples1}
r(\theta) = R [1 + \ve P_n(\cos\theta) ] 
\end{equation}
Fig.~\ref{fig:examples1} (top) illustrates the three surfaces
corresponding to $n = 2,3,4$ that we used in this study.  Varying
$\ve$ from $0$ to $1$, one progressively transforms the sphere of
radius $R$ into three distinct shapes.  A practical advantage of the
choice of Eq.~\eqref{eq:r_examples1} is that the first-order
correction to the total flux vanishes, as
$B_0^\infty = 0$ according to Eq.~\eqref{e:B0}) due to the
orthogonality of Legendre polynomials.  In other words, one has the 
somewhat surprising result that the total flux
onto any surface of the form~\eqref{eq:r_examples1}, with any $n > 0$
and any $0 \leq \ve \leq 1$, is equal to $k_S$, up to the second-order
term $\O(\ve^2)$, and thus $k_{\rm pert} = k_S$.\\
\indent Fig.~\ref{fig:examples1} (bottom) shows the difference between the
total flux as computed numerically and its perturbative approximation
$k_{\rm pert}$, for three surfaces of the kind~\eqref{eq:r_examples1} with $n = 2,3,4$, as a function of
$\ve$.  One can see that $(k - k_{\rm pert})/k_S$ vanishes as $\ve^2$
in the limit $\ve \to 0$, in agreement with our perturbative analysis.
Remarkably, even for $\ve = 1$, which is far beyond the perturbation range, 
the perturbative prediction is within $30 \%$ of the exact value.\\
\indent In order to validate the perturbative approximation for shapes yielding
nontrivial first-order corrections, we considered a second family of
surfaces defined as
\begin{equation}  \label{eq:r_examples2}
r(\theta) = R [1 + \ve P_n^2(\cos\theta)] 
\end{equation}
For these surfaces, one has $B_0^\infty = 1/(2n+1)$, so that 
the non-zero first-order correction reads
\begin{equation} 
\label{eq:k_pert2}
k_{\rm pert}(n) = k_S \biggl(1 +  \frac{\epsilon}{2n+1} \biggr) 
\end{equation}
Fig.~\ref{fig:examples2} shows how the total flux to  
three surfaces of the kind~\eqref{eq:r_examples2} 
corresponding to $n = 2,3,4$ deviates
from our first-order perturbative
approximation $k_{\rm pert}(n)$, eq.~\eqref{eq:k_pert2}.  
Analogously to what reported in Fig.~\ref{fig:examples1}, this
deviation grows as $\ve^2$, confirming that the first-order term was
evaluated correctly.  Remarkably, the deviation at $\ve = 1$ is even smaller for the 
class of surfaces~\eqref{eq:r_examples2}, 
as small as $\simeq 10 \%$.  This lower error (as compared to
Fig.~\ref{fig:examples1}) reflects the fact that the second class of surfaces
defined by Eq.~\eqref{eq:r_examples2} are somewhat {\em rounder}.

%
\section{Conclusions and perspectives}
\label{sec:summary}

\noindent 
In this paper we have tackled the problem of calculating the total
stationary diffusive flux onto partially absorbing surfaces of
revolution.  We have outlined a general perturbative procedure that
holds when the surface of revolution can be described as an
axisymmetric perturbation of a sphere controlled by a single length
scale (see Eq.~\eqref{e:rtheta}).
We worked out an analytical expression for the reaction rate that
agrees surprisingly well with the exact solutions for prolate and
oblate spheroids, even in the case of rather non-spherical shapes with  aspect
ratio up to $0.5$.  The  exact and approximate formulas  for the
reaction rate for partially reactive prolate and oblate spheroids derived in this paper
present their own interest for applications. Furthermore, 
we have tested our perturbative correction by comparing it to 
the exact values of the total flux to different families of 
non-spherical surfaces beyond spheroids. These comparisons 
show that our formula works surprisingly well even far from 
the perturbative limit, $\ve = \mathcal{O}(1)$.   
\indent 
More generally, our perturbative formula~\eqref{e:ratef} for the case of a perfectly
absorbing non-spherical boundary has a simple and general
interpretation.  The total flux to the first order in $\epsilon$ is
the Smoluchowski rate into an equivalent sphere $\S_{\rm eq}$, whose
surface area equals that of the non-spherical surface of revolution
$\S$.  Recalling the definitions~\eqref{e:rtheta} and~\eqref{e:B0},
one has
\begin{eqnarray}
\label{e:eqs}
\S &=& 2\pi \int_0^\pi r(\theta)\sqrt{r^2(\theta) + \left(
                                                               \frac{dr}{d\theta}
                                                             \right)^2 } \sin\theta\,d\theta   \nonumber \\
            &=& 4\pi R^2 \left( 1 + 2 \epsilon B_0^\infty \right) + \mathcal{O}(\epsilon^2)                                                 
\end{eqnarray}
Comparing this result with the perturbative formula~\eqref{e:ratesink}
for the total flux $k$ into the non-spherical sink $\S$, one can
conclude that
\begin{equation}
\label{e:equivrate}
k = 4\pi D R_{\rm eq} c_\infty +   \mathcal{O}(\epsilon^2)
\end{equation}
where $R_{\rm eq} = \sqrt{\S/4\pi}$. We note that the equivalence
expressed by Eq.~\eqref{e:equivrate} proved in this paper had been
conjectured in Ref.~\cite{Berezhkovskii:2007aa} based on intuitive
arguments.  However, our results prove that the equivalence (to
$\mathcal{O}(\epsilon)$ at least) between a non-spherical surface and
a sphere is only limited to perfectly absorbing boundary conditions
(see again formula~\eqref{e:ratef}).  Furthermore, our method can in
principle be employed to investigate whether
formula~\eqref{e:equivrate} is valid beyond the first order in
$\epsilon$.  Moreover, the perturbative treatment presented in this
paper can be also generalized to the case of non-uniform intrinsic
reactivity $\kappa(\theta)$ (see \cite{Grebenkov19c}), which
constitutes an interesting extension for treating diffusion to
non-spherical and non-uniform reactive surfaces.\\
\indent As a final remark, we observe that the mathematics of 
diffusion in the presence of non-spherical reactive boundaries is
important for the description of phoretic locomotion. Relevant
examples include phoretic motion of non-spherical particles whose
surface is only partially reactive (e.g. catalytic), the rest being
chemically inactive~\cite{Uspal2018,Popescu2010}, or autophoresis
purely induced by geometric asymmetries~\cite{Michielin2015}.
Interestingly, in Ref.\citenum{Michielin2015} the authors perform a
perturbative calculation of the autophoretic locomotion of
non-spherical particles that includes computing the stationary
concentration of a solute in the presence of partially absorbing
boundary conditions. Their calculation is formally identical to our
scheme, leading to an expression equivalent to our Eq.~\eqref{e:Bl},
even if their result is cast in the form of an expansion in series of
Legendre polynomials. However, the focus of
Ref.\citenum{Michielin2015} is not the reaction rate itself, which is
the main subject of our analysis, rather it is the resulting phoretic
velocity of the combined diffusion and hydrodynamic problem.

\appendix
\section{Solution for partially absorbing prolate spheroids}
\label{sec:prolate}

Following Ref.~\cite{Traytak:2018ab}, we write the concentration
outside a prolate spheroid of the form $(a,a,b)$ (with semi-axes $a
\leq b$) in prolate spheroidal coordinates $(\alpha,\theta,\phi)$ as
\begin{equation}
u = \sum\limits_{n=0}^\infty A_n \, Q_n(\cosh \alpha) P_n(\cos\theta),
\end{equation}
with unknown coefficients $A_n$.  Here $P_n(z)$ and $Q_n(z)$ are
Legendre polynomials and functions of the first and second kind.  The
Robin boundary condition (\ref{e:SmolBPx2}) reads
\begin{eqnarray}  \label{eq:RBC_prolate}
1 &=&  (u - \Lambda \partial_n u)|_{\pa} = \sum\limits_{n=0}^\infty A_n P_n(\cos\theta) \\  \nonumber
&\times& \biggl[Q_n(\cosh\alpha_0) - \frac{\Lambda \sinh\alpha_0 Q'_n(\cosh\alpha_0)}{a_E \sqrt{\cosh^2 \alpha_0 - \cos^2\theta}} \biggr],
\end{eqnarray}
which should be satisfied by any $\theta \in (0,\pi)$.  Here
$\partial_n = \nabla \cdot \hat{n}$ is the normal derivative
(directed inward), $\Lambda = D/\kappa$ is the reaction length
\cite{Sapoval94,Sapoval02,Grebenkov06,Grebenkov06b,Grebenkov15} and
$\alpha_0$ determines the surface of the prolate spheroid, $\cosh
\alpha_0 = b/a_E$, where $a_E = \sqrt{b^2 - a^2}$.  Multiplying this
condition by $P_m(\cos\theta)\sin\theta$ and integrating over $\theta$
from $0$ to $\pi$, one gets the infinite-dimensional system of linear
equations on coefficients $A_n$
\begin{eqnarray}  \label{eq:system}
\delta_{m,0} &=& \frac{A_m}{2m+1} Q_m(\cosh\alpha_0) - \frac{\Lambda}{2a_E} \\  \nonumber
&\times& \sum\limits_{n=0}^\infty A_n \sinh \alpha_0 Q'_n(\cosh\alpha_0)  F_{m,n}(\cosh\alpha_0),
\end{eqnarray}
where
\begin{equation}
F_{m,n}(z) =  \int\limits_{-1}^1 dx \frac{P_n(x) P_m(x)}{\sqrt{z^2 - x^2}} \,.
\end{equation}
The total diffusive flux onto the prolate spheroid (i.e., the reaction
rate) was computed in Ref. \cite{Traytak:2018ab} as
\begin{equation}  \label{eq:k_prolate}
k = 4\pi D c_\infty a_E A_0 ,
\end{equation}
where $c_\infty$ is the concentration at infinity.  For a perfectly
reactive spheroid ($\Lambda = 0$), the sum in Eq. (\ref{eq:system})
vanishes, and one gets a simple solution for the coefficients $A_m$:
\begin{equation}  \label{eq:Am_prolate0}
A_m = \delta_{m,0} \frac{2m+1}{Q_m(\cosh\alpha_0)} \,.
\end{equation}
From Eqs. (\ref{eq:k_prolate}, \ref{eq:Am_prolate0}), the exact form
of the total flux onto the perfectly reactive prolate spheroid is
recovered \cite{Landau,Berezhkovskii:2007aa}
\begin{equation}  \label{eq:k_prolate0}
\frac{k}{k_S} = \frac{2a_E/b}{\ln \bigl(\frac{1 + a_E/b}{1 - a_E/b}\bigr)} \,,
\end{equation}
where $k_S = 4\pi D c_\infty b$ is the Smoluchowski rate on the sphere
of radius $b$.

In general, however, one needs to truncate the system
(\ref{eq:system}) and then solve it numerically.  Note that the same
solution scheme is valid for Robin boundary condition
(\ref{eq:RBC_prolate}) with a given function $g(\theta)$ on the
left-hand side instead of $1$.  The only difference is that
$\delta_{m,0}$ on the left-hand side of Eq. (\ref{eq:system}) would be
replaced by the integral of $g(\theta)$ with the Legendre polynomial
$P_m(\cos\theta)$.

\subsection{Numerical solution}

For a numerical solution, one needs to truncate the system
(\ref{eq:system}) to some order $\nmax$, to compute the matrix
elements $F_{m,n}(\cosh\alpha_0)$ for $m,n=0,1,\ldots,\nmax$, and then
to invert the resulting matrix of size $(\nmax+1)\times (\nmax+1)$
numerically.  The matrix elements $F_{m,n}(z)$ can be computed exactly
by using the properties of Legendre polynomials.  First, using the
identity
\begin{equation}
P_m(x) P_n(x) = \sum\limits_{k=0}^{\min\{m,n\}} B_{mn}^k \, P_{m+n-2k}(x)
\end{equation}
with
\begin{equation}
B_{mn}^k = \frac{a_k a_{m-k} a_{n-k}}{a_{m+n-k}} \, \frac{2m+2n-4k+1}{2m+2n-2k+1} \,,
\end{equation}
and $a_k = \frac{\Gamma(k+1/2)}{\sqrt{\pi} \Gamma(k+1)}$ (with $a_0 =
1$), we get for even $m+n$
\begin{equation}
F_{m,n}(z) = \sum\limits_{k=0}^{\min\{m,n\}} B_{mn}^k \, V_{(m+n)/2-k}(z),
\end{equation}
where
\begin{equation}
V_n(z) = \int\limits_{-1}^1 dx \frac{P_{2n}(x)}{\sqrt{z^2 - x^2}} \,,
\end{equation}
while $F_{m,n}(z) = 0$ for odd $m+n$ due to the symmetry of Legendre
polynomials.  Second, the above integral can be computed via recursive
relations:
\begin{eqnarray} \nonumber
V_n(z) &=& \frac{(4n-3)[(4n-1)(4n-5)z^2 - 2(4n^2-6n+1)]}{(2n)^2(4n-5)}  \\
&\times& V_{n-1}(z) - \frac{(4n-1)(2n-3)^2}{(4n-5)(2n)^2} V_{n-2}(z) \,.
\end{eqnarray}
with 
\begin{eqnarray*}
V_0(z) &=& 2\arcsin(1/z) \\
V_1(z) &=& \biggl(\frac32 z^2 - 1 \biggr) \arcsin(1/z) - \frac32 \sqrt{z^2-1} \,.
\end{eqnarray*}
As $F_{m,n}$ vanishes for odd $m+n$, the linear equations in the
system (\ref{eq:system}) decouple into two subsystems, one for the
coefficients $A_{2m}$ and the other for the coefficients $A_{2m+1}$.
Moreover, as $\delta_{m,0}$ in the left-hand side of the system is not
zero only for $m=0$, all the odd coefficients $A_{2m+1}$ are zero.  In
turn, the system for even coefficients $A_{2m}$ can be written as
\begin{eqnarray}  \label{eq:system_prolate2}
\delta_{m,0} &=& \frac{A_{2m}}{4m+1} Q_{2m}(\cosh\alpha_0)   \\  \nonumber
&-& \frac{\Lambda}{2a_E} \sum\limits_{n=0}^\infty F_{2m,2n}(\cosh\alpha_0)  
    \sinh \alpha_0 Q'_{2n}(\cosh\alpha_0) A_{2n}   \,.
\end{eqnarray}
Truncating the sum to $n = \nmax$, computing
$F_{2m,2n}(\cosh\alpha_0)$ and solving numerically the truncated
system with $\nmax+1$ equations onto the coefficients $A_0, A_2,
\ldots, A_{2\nmax}$, one computes the reaction rate $k$ from
Eq. (\ref{eq:k_prolate}).

\subsection{Approximate solution}

Solving the truncated system (\ref{eq:system_prolate2}) for several
values of $\nmax$, we checked that the coefficient $A_0$, determining
the reaction rate, can be accurately computed from low-order
truncations.  In other words, the estimated $A_0$ converges very
rapidly to its limit as $\nmax$ increases.  As illustrated in
Fig. \ref{fig:prol_obl_k}, even the lowest-order truncation with
$\nmax = 0$ is accurate over the whole range of reactiony lengths
$\Lambda$ even when the minor semi-axis $a$ is as small as $b/10$.
This zeroth-order approximation yields
\begin{equation}  \label{eq:A0_prolate_n0}
\frac{k}{k_S} \simeq \frac{a_E}{b}
\biggl(\frac12 \ln \biggl(\frac{1+a_E/b}{1-a_E/b}\biggr) + \frac{\Lambda}{b} \frac{\arcsin(a_E/b)}{(1 - a_E^2/b^2)^{1/2}}\biggr)^{-1} ,
\end{equation}
where $k_S = 4\pi D c_\infty b$ is the Smoluchowski rate for a sphere
of radius $b$.  For a perfectly reactive spheroid,
Eq. (\ref{eq:A0_prolate_n0}) at $\Lambda = 0$ yields the exact result
(\ref{eq:k_prolate0}).  As expected, one retrieves the Collins-Kimball
result for a partially reactive sphere in the limit $a \to b$:
\begin{equation} \label{eq:CK}
\frac{k}{k_S} = \biggl(1 + \frac{D}{\kappa b}\biggr)^{-1} .
\end{equation}
The lowest-order approximation becomes less accurate in the limit
$a\to 0$ when prolate spheroids are getting closer to the shape of a
needle.  In this case, one can consider the first-order truncation
with $\nmax = 1$, in which case the truncated $2\times 2$ system of
equations can be again solved explicitly, yielding a cumbersome but
remarkably accurate solution.\\
\indent Figure \ref{fig:prol_obl_k} illustrates the behavior of the scaled
reaction rate $k/k_S$ for a partially reactive prolate spheroid as a
function of $D/(\kappa b) = \Lambda/b$.  For all considered values of
$\Lambda/b$ and $a/b$, the numerical solution of the truncated system
of linear equations converges very rapidly with the truncated order
$n_{\rm max}$.  In particular, one can see that solutions with $n_{\rm
max} = 5$ (referred to as ``exact solution'') and $n_{\rm max} = 1$
are barely distinguishable over all $\Lambda/b$ even for a very narrow
spheroid with $a/b = 0.01$.  This observation confirms that this
explicit approximate solution is very accurate.  A simpler
zeroth-order approximate solution from Eq. (\ref{eq:A0_prolate_n0}) is
also very accurate, in spite of small deviations at large reactivity
(small $\Lambda/b$).  Finally, the perturbative solution
(\ref{e:ratefex}) is accurate for small and moderate $\epsilon = 1 -
a/b$, and fails only when $\epsilon$ becomes close to $1$ (i.e., $a$
close to $0$).

\subsection{Perturbative solution}

An approximate solution of the above system of equations can be
obtained in the limit when the prolate spheroid approaches a sphere,
i.e., $a = b(1 - \epsilon)$, so that $a_E^2 = 2b^2 \epsilon +
\O(\epsilon^2)$, and thus $\cosh\alpha_0 = 1/\sqrt{2\epsilon} \to
\infty$.  In the limit $z\to\infty$, we have thus
\begin{equation}
F_{m,n}(\cosh\alpha_0) = 2\sqrt{2\epsilon} \biggl(1 + f_{m,n} \epsilon + \O(\epsilon^{2})\biggr),
\end{equation}
with
\begin{equation}
f_{m,n} = \frac12 \int\limits_{-1}^1 dx\, x^2 \, P_m(x) \, P_n(x)
\end{equation}
(this integral can be evaluated explicitly).
Denoting
\begin{equation}
a_n = A_n \biggl(Q_n(\cosh\alpha_0) - \frac{\Lambda}{b} \sinh \alpha_0 Q'_n(\cosh\alpha_0)\biggr),
\end{equation}
we rewrite the system (\ref{eq:system_prolate2}) up to the first order
in $\epsilon$ as
\begin{equation}  \label{eq:system2}
\delta_{m,0} = \frac{a_{2m}}{4m+1} - \epsilon \sum\limits_{n=0}^\infty f_{2m,2n} a_{2n} e_{2n}  + \O(\epsilon^2),
\end{equation}
where
\begin{equation}
e_n = \lim\limits_{\epsilon\to0} \frac{(\Lambda/b) \sinh \alpha_0 Q'_n(\cosh\alpha_0)}{Q_n(\cosh\alpha_0) 
- (\Lambda/b) \sinh \alpha_0 Q'_n(\cosh\alpha_0)} \,.
\end{equation}
%
One gets thus from Eq. (\ref{eq:system2}) with $m = 0$:
\begin{equation}
a_0 = \frac{1 + \epsilon e_2 f_{0,2} a_2}{1 - \epsilon e_0 f_{0,0}} \,,
\end{equation}
whereas $a_2 = \O(\epsilon)$ so that the second term in the numerator
can be neglected, yielding (with $f_{0,0} = 1/3$)
\begin{equation}
a_0 = 1 + \frac{\epsilon}{3} e_0 + \O(\epsilon^2).
\end{equation}
Given that $Q_0(z) = \frac12 \ln \frac{z+1}{z-1}$ and $Q'_0(z) =
\frac{1}{1-z^2}$, we obtain $e_0 = - 1/(1+h)$, where $h = b/\Lambda =
\kappa b/D$.  We get thus
\begin{equation}
A_0 = \frac{(b/a_E) h}{1+h} \biggl(1 - \epsilon \frac{2(h+2)}{3(h+1)} + \O(\epsilon^2)\biggr),
\end{equation}
from which
\begin{equation}
\label{e:ratefex}
\frac{k}{k_S} = \frac{h}{1+h} - \epsilon \frac{2h(h+2)}{3(h+1)^2} + \O(\epsilon^2).
\end{equation}
This expression agrees with our general formula~\eqref{e:ratef} for
$f(\theta) = - \sin^2\theta$ that approximately describes a prolate
spheroid for small $\epsilon$.

\section{Oblate spheroid}
\label{sec:oblate}

As computations for an oblate spheroid of the form $(b,b,a)$ (with
semi-axes $a \leq b$) are similar, we only sketch the main steps.  In
the oblate spheroidal coordinates $(\alpha,\theta,\phi)$, an
axiosymmetric solution of the Laplace equation outside the spheroid
reads
\begin{equation}
u = \sum\limits_{n=0}^\infty A_n \, Q_n(i\sinh \alpha) \, P_n(\cos\theta),
\end{equation}
where the coefficients $A_n$ are determined from the Robin boundary
condition
\begin{eqnarray*}
1 &=& (u - \Lambda \partial_n u)|_{\pa} = \sum\limits_{n=0}^\infty A_n P_n(\cos\theta) \\
&\times& \biggl(Q_n(i\sinh \alpha_0) - \frac{i \Lambda \cosh\alpha_0 Q'_n(i\sinh\alpha_0)}{a_E \sqrt{\cosh^2\alpha_0 - \cos^2\theta}} \biggr),
\end{eqnarray*}
from which
\begin{eqnarray}
\delta_{m,0} &=& \frac{A_m}{2m+1} Q_m(i\sinh\alpha_0) - \frac{i\Lambda}{2a_E} \\  \nonumber
&\times& \sum\limits_{n=0}^\infty A_n \cosh \alpha_0 Q'_n(i\sinh\alpha_0) F_{m,n}(\cosh\alpha_0).
\end{eqnarray}
Note that the boundary of the oblate spheroid is defined again as
$\cosh\alpha_0 = b/a_E$, with $a_E = \sqrt{b^2 - a^2}$.  The total
diffusive flux reads as (see Ref.~\cite{Grebenkov:2018aa} for
technical details)%
\footnote{
We note that Appendix C of Ref.~\cite{Grebenkov:2018aa} contains
several technical misprints that do not alter the final results.}
\begin{equation}  \label{eq:rate_oblate}
k = 4\pi D c_\infty a_E A_0/i .
\end{equation}
As previously, only the even coefficients matter so that 
\begin{eqnarray}  \label{eq:system_oblate}
\delta_{m,0} &=& \frac{A_{2m}}{4m+1} Q_{2m}(i\sinh\alpha_0) - \frac{i\Lambda}{2a_E} \\  \nonumber
&\times& \sum\limits_{n=0}^\infty A_{2n} \cosh \alpha_0 Q'_{2n}(i\sinh\alpha_0) F_{2m,2n}(\cosh\alpha_0).
\end{eqnarray}

For a perfectly reactive oblate spheroid ($\Lambda = 0$), the sum
vanishes, so that $A_0 = 1/Q_0(i\sinh\alpha_0)$, from which
Eq. (\ref{eq:rate_oblate}) helps to retrieve the exact formula for the
reaction rate \cite{Landau,Berezhkovskii:2007aa}
\begin{equation}
\frac{k}{k_S} = \frac{a_E/b}{\arcsin(a_E/b)} \,,
\end{equation}
where we used $Q_0(i\sinh\alpha_0) = -i \arcsin(1/\cosh\alpha_0)$.  

For a partially reactive spheroid, the zeroth-order truncation of the
system (\ref{eq:system_oblate}) yields the following approximation for
the reaction rate
\begin{equation} \label{eq:A0_oblate_n0}
\frac{k}{k_S} \simeq \frac{a_E/b}{\arcsin(a_E/b)} \, \frac{1}{1 + \Lambda/b} \,.
\end{equation}
In the limit $a\to b$, one retrieves the Collins-Kimball result
(\ref{eq:CK}) for a partially reactive sphere.  In the opposite limit
$a \to 0$, one finds the reaction rate for a partially absorbing disk
of radius $b$:
\begin{equation}
\frac{k}{k_S} \simeq \frac{2/\pi}{1+\Lambda/b} \,.
\end{equation}
The accuracy of this approximation is illustrated in
Fig. \ref{fig:prol_obl_k}.  An even higher accuracy can be achieved by
using the truncation order $\nmax = 1$, for which the resulting
$2\times 2$ matrix can be solved explicitly.  In comparison to
Eq. (\ref{eq:A0_prolate_n0}) for the prolate spheroid, one can note
that the geometric aspect is fully disentangled from the reactive
aspect, which is accounted via the factor $1/(1 + \Lambda/b)$, as for
the sphere.\\
\indent Skipping technical detials, we also provide the perturbative solution
of Eqs. (\ref{eq:system_oblate}) up to the first order in $\epsilon$:
\begin{equation} \label{eq:k_pert_oblate}
\frac{k}{k_S} = \frac{h}{1+h} - \epsilon \frac{h(h+2)}{3(h+1)^2} + O(\epsilon^2).
\end{equation}
This solution agrees with our general formula~\eqref{e:ratef} for
$f(\theta) = - \cos^2\theta$ that approximately describes an oblate
spheroid for small $\epsilon$.

\section{Validation by a finite-element method}
\label{sec:A_FEM}

We seek a numerical solution of the boundary value problem~(\ref{e:SmolBPx})
in the limit $h\to\infty$.
%
For this purpose, we use a finite-element method implemented in Matlab
PDEtool.  As this method meshes the computational domain, the latter
should be bounded by an artificially imposed outer boundary
$\Gamma_\alpha$.  For convenience, we choose $\Gamma_\alpha$ to be a
sphere of radius $R_\alpha = R/\alpha$ centered at the origin:
$\Gamma_\alpha = \{\vec r\in\R^3 ~:~ \| \vec r\| = R_\alpha\}$.  We
aim thus to solve numerically the modified problem
\begin{subequations}  \label{eq:PDE_modified}
\begin{align}  \label{eq:Laplace}
\nabla^2 u_\alpha & = 0   \\  \label{eq:Robin}
(1 - u_\alpha)|_{\pa'} & = 0, \\  \label{eq:BC_Gamma}
(\nabla u_\alpha \cdot \hat{n})|_{\Gamma_\alpha} - \frac{1}{R_\alpha}\, u_\alpha|_{\Gamma_\alpha} & = 0, 
\end{align}
\end{subequations}
in the new bounded domain: $\Omega_\alpha = (\R^3 \setminus T') \cap
\{\vec r\in\R^3 ~:~ \| \vec r\| < R_\alpha\}$.  As $\alpha$ goes to
zero, the outer boundary $\Gamma_\alpha$ moves away from the target
$T'$, $\Omega_\alpha$ blows up and approches $(\R^3 \setminus T')$,
and $u_\alpha$ converges to the solution $u$ of the original problem
(\ref{e:SmolBPx}).  Even though the boundary condition at the imposed
outer boundary $\Gamma_\alpha$ does not matter in the limit $\alpha
\to 0$, we choose the Robin condition~(\ref{eq:BC_Gamma}) to speed up
the convergence.  In fact, if the surface $\pa'$ was a sphere of
radius $1$, then the solution of Eqs. (\ref{e:SmolBPx}) would simply
be $1/r$. 
Moreover, for an arbitrary surface $\pa'$, the solution of
Eqs. (\ref{e:SmolBPx}) is expected to asymptotically decay as
$\gamma/r$ (with some constant $\gamma$), and the Robin condition
(\ref{eq:BC_Gamma}) is again consistent with this behavior.  In
summary, even so the condition (\ref{eq:BC_Gamma}) could be replaced
by more common Dirichlet or Neumann conditions, the choice
(\ref{eq:BC_Gamma}) is preferred.\\
\indent The axial symmetry of the domain $\Omega_\alpha$ reduces the original
three-dimensional problem to a two-dimensional one.  The Laplace
equation (\ref{eq:Laplace}) can be written in a standard matrix form
in spherical coordinates $(r,\theta,\phi)$ as
%
\begin{equation}
- \left(\begin{array}{c} \partial_r \\ \partial_\theta \end{array}\right)^\dagger c 
\left(\begin{array}{c} \partial_r \\ \partial_\theta \end{array}\right) u = 0
\end{equation}
where $c$ is the diagonal $2\times 2$ matrix with the diagonal
elements equal to $r^2 \sin\theta$ and $\sin\theta$.  The
two-dimensional computational domain is then:
\begin{equation}
C_\alpha = \{ (\theta,r) \in \R^2 ~:~ 0 < \theta < \pi , ~ r(\theta) < r < R_\alpha \} ,
\end{equation}
i.e., it is a kind of rectangle of width $\pi$, whose bottom ``edge''
is curved and determined by a given function $r(\theta)$.  The
boundary conditions (\ref{eq:Robin}, \ref{eq:BC_Gamma}) are set on the
bottom and top ``edges'', whereas the Neumann condition
$\partial_\theta u_\alpha = 0$ is imposed on the left and right edges.\\
\indent Once the problem (\ref{eq:PDE_modified}) is solved numerically, the
total diffusive flux follows as
\begin{equation}
k_\alpha = - c_\infty D \int\limits_{\pa'} dS \, (\nabla u_\alpha \cdot \hat{n}).
\end{equation}
%
%
%
%
Alternatively, the flux conservation allows one to compute the total
flux by integrating over the outer boundary $\Gamma_\alpha$ and using
Eq. (\ref{eq:BC_Gamma}),
\begin{equation}
k_\alpha  = 2\pi c_\infty D R_\alpha \int\limits_0^\pi d\theta \, \sin\theta \,  u_\alpha(\theta,R_\alpha),
\end{equation}
which provides a much simpler way to compute the total flux. 
As $k_\alpha$ approaches $k$ of the original problem in the
limit $\alpha \to 0$, this expression can be used as an approximation
of $k$ for $\alpha$ small enough.\\
\indent For each surface determined by a given function $r(\theta)$, we
computed the total flux $k_\alpha$ by setting $R_\alpha = 10 R$ and
the maximal mesh size $\eta_{\rm max}$ of $0.05$.  To check the
accuracy of numerical computations, $k_\alpha$ was also computed by
using either $R_\alpha = 20 R$ (farther outer boundary), or $\eta_{\rm
max} = 0.025$ (finer mesh).  For all considered surfaces, these
changes did not almost affect $k_\alpha$ (the relative error being
much smaller than $1\%$).  This observation confirms the high accuracy
of our numerical computation.

\section*{Acknowledgements}
FP would like to thank Stefano Angioletti-Uberti for enlightening discussions.



\balance



\end{document}